\documentclass[a4paper,11pt]{article}
\usepackage{amssymb}
\usepackage{paralist}
\usepackage{graphicx}
\usepackage{textcomp}
\usepackage{xcolor}
\usepackage{glossaries}
\usepackage{hyperref}
\usepackage[font=small, labelfont=bf]{caption}
\usepackage[utf8]{inputenc}
\usepackage[export]{adjustbox}
\newacronym{CI}{CI}{Continuous Integration}
\newacronym{CICD}{CI/CD}{Continuous Integration / Continuous Deployment}
\newacronym{CD}{CD}{Continuous Deployment}
\newacronym{COVID-19}{COVID-19}{Coronavirus disease 2019}
\newacronym{CPAP}{CPAP}{Continuous positive airway pressure}
\newacronym{CSV}{CSV}{Comma-separated value}
\newacronym{DES}{DES}{discrete event simulation}
\newacronym{DIVI}{DIVI}{Deutsche interdisziplinäre Vereinigung für Intensiv- und Notfallmedizin}
\newacronym{EC}{EC}{Evolutionary Computation}
\newacronym{ECiP}{ECiP}{Evolutionary Computation in Practice}
\newacronym{ESPOL}{ESPOL}{Escuela Superior Politécnica del Litoral}
\newacronym{EGO}{EGO}{Efficient Global Optimization}
\newacronym{ETL}{ETL}{Extract, Transform, and Load}
\newacronym{ICU}{ICU}{Intensive Care Unit}
\newacronym{ISRES}{ISRES}{Improved Stochastic Ranking Evolution Strategy}
\newacronym{LHD}{LHD}{Latin Hypercube Design}
\newacronym{R}{R}{R language and environment for statistical computing and graphics}
\newacronym{RKI}{RKI}{Robert Koch Institut}
\newacronym{RMSE}{RMSE}{Root Mean Squared Error}
\newacronym{RSM}{RSM}{Response surface methodology}
\newacronym{simmer}{simmer}{``Discrete event simulation package for R''}
\newacronym{SHERPA}{SHERPA}{Simultaneous Hybrid Exploration that is Robust, Progressive and Adaptive}
\newacronym{SMBO}{SMBO}{Surrogate Model-Based Optimization}
\newacronym{SMBSA}{SMBSA}{Surrogate Model-Based Sensitivity Analysis}
\newacronym{SPOT}{SPOT}{Sequential Parameter Optimization Toolbox}
\newacronym{PSSO}{PSSO}{Prüfungs- und Studierendenservice Online}
\newacronym{STEM}{STEM}{Science, Technology, Engineering, and Mathematics}
\newacronym{DZHW}{DZHW}{Deutsches Zentrum für Hochschul- und Wissenschaftsforschung GmbH}
\newacronym{FH}{FH}{Fachhochschule}
\newacronym{TH}{TH Köln}{Technische Hochschule Köln}
\newacronym{GitLab}{GitLab}{GitLab}
\newacronym{J2EE}{J2EE}{Java 2 Platform, Enterprise Edition}
\newacronym{GPA}{GPA}{Grade Point Average}
\newacronym{GUI}{GUI}{Graphical User Interface}
\newacronym{CSS}{CSS}{Cascading Style Sheets}
\newacronym{CRAN}{CRAN}{The Comprehensive R Archive Network}
\newacronym{HTML}{HTML}{Hypertext Markup Language}

\begin{document}
\title{\textbf{An approach to optimize study programs using Discrete Event Simulation}}

\author{
    \textbf{Marcel Dröscher}\\
    \textit{Institute for Data Science, Engineering,}\\
    \textit{and Analytics - TH Köln}\\
    Cologne, Germany\\
    marcel\_ulrich.droescher@smail.th-koeln.de
  \and
    \textbf{Alpar Gür}\\
    \textit{Institute for Data Science, Engineering,}\\
    \textit{and Analytics - TH Köln}\\
    Cologne, Germany\\
    alpar.guer@smail.th-koeln.de\\
    \and
    \textbf{Nicolas Rehbach}\\
    \textit{Institute for Data Science, Engineering,}\\
    \textit{and Analytics - TH Köln}\\
    Cologne, Germany\\
    nicolas\_alexander.rehbach@smail.th-koeln.de
}

\author{
\textbf{Marcel Dröscher}, \textbf{Alpar Gür}, \textbf{Nicolas Rehbach}\\
\textit{Institute for Data Science, Engineering, and Analytics}\\
    \textit{TH Köln}\\
    \underline{marcel.droescher@t-online.de}\\
    \underline{guralpar@hotmail.com }\\
    \underline{rehbach.nicolas@gmail.com}\\
}

\date{\today}
\maketitle

\begin{abstract}
\textbf{Creating a study program for optimal academic completion is a complex assignment. Especially programs in the science, technology, engineering, and mathematics field are known for extended completion time as well as high drop-out rates throughout the years.\\
Drop-outs are caused by various reasons and can not be directly generalized. This leads to unnecessary costs for the students and the university. Reasons for dropping out of university could be students failing classes too often or poorly designed study programs causing a loss of interest in a subject. Besides dropping out of university, students are often having trouble with specific classes. This results in postponing certain classes, which causes a bottleneck in the overall progression and delays graduation. \\
To achieve a better understanding of a general student’s progression as well as finding mentioned bottlenecks in an average academic progression a discrete event simulation was created. The insights gained by the simulation shall furthermore be used by Technische Hochschule Köln to analyze mentioned problems, find solutions and create a healthier environment for optimal academic progression.}
\end{abstract}

\small	
  \textbf{Keywords: \textit{Discrete event simulation; R; extract, transform, and load; \\
    student flow; Simmer; Shiny; academic progress; \gls{TH}}} \newpage

\section{Introduction}
Students dropping out of university is a common problem in every country. It is complicated to counteract against students leaving the university since the reasons vary and are usually unknown. An analysis of students withdrawing from their academic completion shows that loss of motivation and missing identification with their future occupation were key reasons for students studying in humanities \cite{Heublein}. \\

According to the study made by the \Gls{DZHW} study programs in \gls{STEM} have the highest drop-out numbers out of all study fields in Germany with an average of 41 percent of drop-outs for freshmen starting in 2012/2013. While there are several kinds of programs falling into \gls{STEM} it is important to distinguish between each study program. Typically, Mathematics, Chemics, Physics, and Geoscience have a drop-out rate of over 40 percent with math having the highest drop-out rate of 54 percent. On the other hand, programs like Biology and Geography have respectively lower drop-out rates at 28 and 19 percent.\\

Faculty 10 of \gls{TH} offers bachelor and master degrees in the fields of Computer Science and Engineering where high drop-out rates and prolonged academic progress are a common problem. \\
Therefore, research was started to analyze how the organisation could reduce those numbers by planning better study programs and finding problematic areas within the academic completion of their students. \\

This article describes a first implementation of a simulation using \Gls{simmer}, tracking the academic progression of 1408 students throughout 28 courses. The simulation takes several parameters into account considering the probability of students choosing a course, passing or failing a course, having the prerequisites to apply for a course, and more. Taking these values into account, the simulation generates data considering students' drop-out rates, their overall study duration as well as course occupancy rates. While the amount of students is the actual number of students enrolled in the Wirtschaftsinformatik (eng. Business Information studies) study program of \gls{TH} from the year 2015 to 2020, the possibilities to fail a course were randomized. For future development using real student data is crucial to gain an actual overview of the students’ academic progression inside their study program. \\

The paper is structured as follows: after the introduction, a brief overview of used methods and tools is being given. Part two includes alternative approaches and relevant research papers, providing first ideas for structuring data and planning the simulation.  
The next section will grant a deeper insight into the simulation including data acquisition and CI/CD, the \gls{DES}, and visualizing the results by using an interactive Shiny app.
In the last section, the results are analyzed and a discussion about future work and real-world adaptation is being held. 

\section{Methods}
The simulation is modelled for the projection of different student groups' academic progress over the course of several semesters. \Gls{DES} have “a great potential to help the understanding of the nature of these programs, allowing educational authorities to evaluate the efficiency of current academic systems and predict the performance of new curricular designs” \cite{8120908}. Therefore, the \gls{DES} is a good match for the use case, being the reason for choosing it over the Markov chain model or a time-series analysis. Real data has been utilized to determine the course occupancy rate, however the rest of the parameter values are initialized heuristically. \\

The simulation serves as an engine, meaning whenever the real data is accessible, it can be plugged in to get more accurate results. It is compatible for different domains with the same use case. In order to sustain the full functionality and data integrity, an \gls{ETL} pipeline has been implemented on a GitLab server which also contains all project files. The simulator has been programmed using \gls{R} and the results are visualized with Shiny \cite{shiny} and ggplot2 \cite{ggplot2}.\\

\section{Related Work}
\label{sec:rel_work}
\subsection{The development of dropout rates at German universities}
To gain information about the actual drop-out rates and when a student officially counts as a drop-out, a deeper look into studies is required. Therefore, a specific overview of a study presented by the \gls{DZHW} is given \cite{Heublein}. \\

After giving a short introduction to the topic, the study gives insight into the drop-out ratios in German universities divided into bachelor, master, and state examination students as well as foreign students studying in Germany. In the end, an overview of the methodology used to calculate drop-out rates is given. \\
The study includes information about which student counts as a drop-out, the differences between a university and a technical university and how drop-out rates are calculated in Germany. For reproducibility, a deeper look into the drop-out terms of each country should be held. \\

In Germany, students count as a drop-out who have matriculated at a German university or technical university but left without obtaining their first graduation. If a student decides to switch universities, matriculates in a different study field, or pursues a second bachelor degree he will not be counted as a drop-out. Moreover, master students dropping out of their program will also not be taken into the overall account of drop-outs, since they have already officially obtained a bachelor's degree. It is still worth noting, that the percentage of master students leaving their program is significantly lower than the percentage of bachelor students dropping out. Therefore, the main selection of students pursuing their academic careers will happen during bachelor programs. \\

Another interesting aspect the study found out is that universities usually have a higher number of drop-outs than in a \gls{FH}. While both university types offer the same degree there are still differences in the scientific orientation. Usually, \glspl{FH} are more interested in application-oriented basic research while universities are looking for theoretical approaches \cite{SchwindsacklMartina2011UvF:}.  \\

Another important aspect are foreign students. The study found out that there is a large gap between drop-outs who grew up in Germany and students coming from another country to study in Germany. However, it is important to mention that these drop-out rates only count for an unsuccessful academic progression at one university in Germany. Each student who started studying in Germany and will finish in his/her home country will therefore be counted as a drop-out in Germany. Accordingly, one can assume that these rates are usually overestimated.\\

For calculating students’ drop-out rates the study uses a cohort comparison grouping students corresponding to their start of studying. Since graduates of the examined year are usually from different cohorts, the selected cohort will be compared to all relevant cohorts finishing in the examined year. \\
Because of changing numbers of students applying and changing study duration, the study applies several relevant steps to correct these numbers. First, adjusting the number of graduating students in comparison to growing or sinking number of first-year students as well as adjusting the number of graduating students concerning changing study duration. Furthermore, first-year students who already finished a bachelor's degree will not be considered. Lastly, graduates will be calculated back to their original university type, degree, and subject group in which they were enrolled when starting as first-year students.

\subsection{Discrete event simulation for student flow in academic study periods}
\label{subsec:desrw}
A study by Angel Fiallos and Xavier Ochoa published in 2017 investigates a problem very similar to the one discussed in this work \cite{8120908}. Whilst there have been works on predicting student movement and performance in educational programs using Markov models \cite{markovhs}, this study takes a different approach by using \gls{DES} under the \gls{J2EE} platform to simulate the student flow at the computer science bachelor's program of \gls{ESPOL}. The reasoning being "Discrete Event Simulation is more 
adaptable to common, practical real-world simulations and it 
more easily accommodates the complexities and 
interdependencies of the many components involved in most 
systems of interest."\cite{8120908} The simulation considered six input variables:
\begin{itemize}
    \item[\texttt{X1}:] Number of students to be enrolled in the simulated 
academic periods. 
    \item[\texttt{X2}:] Historical \gls{GPA} by courses and groups of students.
    \item[\texttt{X3}:] Student record history by semester.
    \item[\texttt{X4}:] Historical dropout rates per student group 
    \item[\texttt{X5}:] Curriculum grid and courses 
    \item[\texttt{X6}:] Course prioritization coefficients 
\end{itemize}
and provided 2 output variables:
\begin{itemize}
    \item[\texttt{Y1}:] Career completion time per student
    \item[\texttt{Y2}:] Academic records per semester and student's GPA. 
\end{itemize}
The results display a high degree of accuracy in predicting the behaviour of students in picking and passing courses. The study's prototype showed that \gls{DES} can be a valuable tool to predict student behaviour in educational programs to allow administrations to know beforehand where possible bottlenecks could emerge. It can also be used to investigate possible effects of new policies or changes in the curriculum before deploying them. This study has been the most influential on this work. Especially, the solution for the course selection process has been adapted only in slight variation.

\subsection{Markov Analysis of Students’ Performance and Academic Progress in Higher Education}
\label{subsec:maacprog}
Baggia et. al. \cite{markovprog} investigate the subject with a similar approach using absorbing Markov chain to analyze the pattern of students' enrolment and their academic performance in a Slovenian higher education institution. The paper's objective is to monitor quality and effectiveness of a given higher education study program and to predict the students enrolment for the next three academic years.\\

The model is defined according to seven different possible stages, from a local education system. The key indicator of student performance is represented as the "fraction of students who succeed to progress to a higher stage during one academic year". Moreover, it provides an expected study duration from enrolment to graduation considering eight consecutive academic seasons.

The model had been applied in a real-world case at the University of Maribor, Faculty of Organizational Sciences and delivered useful insights for the organisation in order to improve their policies.

\section{Study}
\subsection{Extract, Transform, and Load}
Data used for the simulation consists of pseudonymized student data extracted directly from the \gls{PSSO} hosted by \gls{TH}, which is the main database for student grading and student service. \\ 

Therefore, an \gls{ETL} process is used to remodel students' data. The main idea of an \gls{ETL} process is to undergo each part of its name and transform used data into a consistent and correct form for further usage. The “extract” phase describes the data procurement from several sources oftentimes being data warehouses or files. Inside the “transform” phase the extracted data will be cleaned "which consists of detecting and correcting the mentioned problems. In particular, this phase includes data imputation that detects and corrects missing values“ \cite{ribeiro}. The final phase “load” will be triggered once the data is in its desired form. Now data can be loaded into the data storage of choice for further analysis. 
Several aspects were analyzed to gain an insight into specific student groups' performance and course occupancy.\\

\begin{figure}[ht]
	\centering
	\small
  \includegraphics[width=1\linewidth]{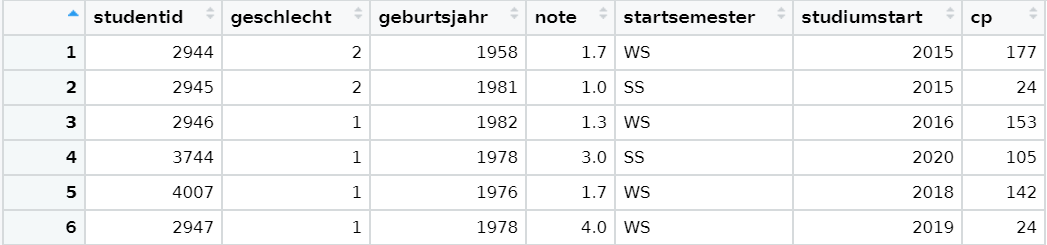}
	\caption{\texttt{A data set containing general students' data. This data is used inside the simulation for several analysis aspects in view of a student's academic success.}}
	\label{figure:studentsdata}
\end{figure}

The data processing part consists of five functions that prepare the data for the simulation. First, a student data frame will be created by using a \gls{CSV} file provided by \gls{TH}. This file contains a randomized student number, the student’s year of birth, their gender, their start of the study as well as their current residence. Because of missing grades, every student received a random average grade (see \textbf{Figure \ref{figure:studentsdata}}).\\

\begin{figure}[ht]
	\centering
	\small
  \includegraphics[width=1\linewidth]{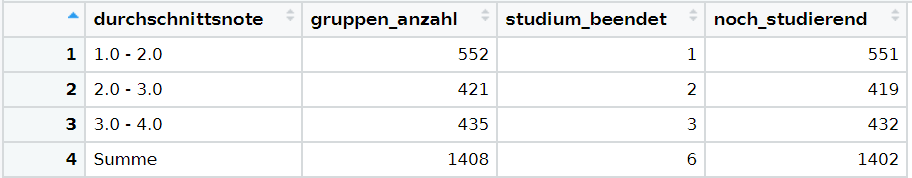}
	\caption{\texttt{A data set containing students grouped by their grading. The grouped data gives a basic overview about the students average success rates concerning their study program.}}
	\label{figure:studentgroups}
\end{figure}

Secondly, students are grouped into three groups by their mean grades ranging from 1.0 to 2.0 (US: A and B), 2.0 to 3.0 (US: B and C), and 3.0 to 4.0 (US: C and D) giving a rough estimate of the number of students in each group and the number of finished degrees as seen in \textbf{Figure \ref{figure:studentgroups}}. \\

\begin{figure}[ht]
	\centering
	\small
  \includegraphics[width=1\linewidth]{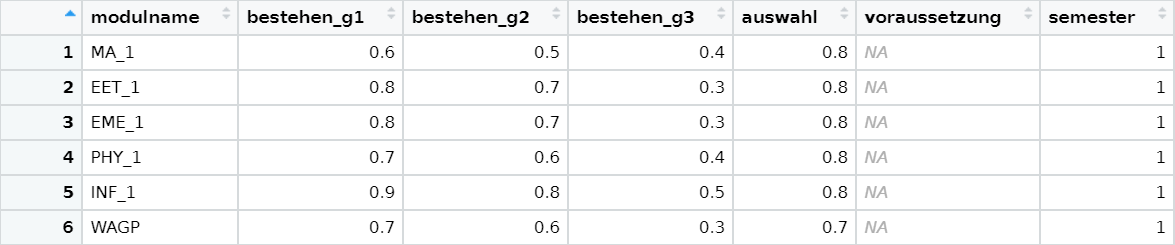}
	\caption{\texttt{A data set containing information about a groups success rate, the probability of choosing the module and prerequisites. This information is used inside the simulation to calculate each students' success rate in consideration of his group.}}
	\label{figure:modules}
\end{figure}

Thirdly, a data frame containing the course of a study program was created. An example can be seen in \textbf{Figure \ref{figure:modules}}. This data frame contains the average success rate of each student group. A group's probability to pass a course is described by the variables bestehen\textunderscore g1, bestehen\textunderscore g2 and bestehen\textunderscore g3. Furthermore, the probability of a student choosing a certain course, requirements to participate in a course, and the recommended semester for each course are given.\\

Moreover, it is important to specify the number of simulated students. Therefore, a fourth function was developed to filter out all the students, that are not relevant for the analyzed period. Considering the students' entry in the simulation a column with the entry semester of each student was added. This is needed to know in which year a student enters and leaves the simulation.\\
The last function adds a semester to each student, used for students still inside the simulation who have not finished or dropped out of university yet. \\

Overall, this data is initialized for different parameters inside the simulation. Examples could be estimating the occupancy rate of a course, the study duration for each student considering his average grade, or the drop-out rate for each student group. \\

\begin{table}[h]
\begin{tabular}{|l|c|r|}
\hline
\textbf{Variable name} & \textbf{Description} \\
\hline
studentid & a unique identifying number for each student \\
geschlecht & a student's gender \\
note & a student's grade\\
startsemester & a student's enrollment semester\\
studiumstart & a student's enrollment year\\
semestercount & a counter variable for the simulation\\
durchschnittsnote & grading range of a students group\\
gruppen\textunderscore anzahl & amount of students in each groups\\
studium\textunderscore beendet & amount of students finished studying\\
noch\textunderscore studierend & amount of students still studying\\
modulname & name of a certain module\\
bestehen\textunderscore g1 & probability of group 1 to pass the exam\\
bestehen\textunderscore g2 & probability of group 2 to pass the exam\\
bestehen\textunderscore g3 & probability of group 3 to pass the exam\\
auswahl & probability of choosing a module\\
voraussetzung & necessary prerequisite\\
semester & suggested semester to participate in the module\\
\hline
\end{tabular}
\caption{Description of used variables inside the simulation}
\label{variables}
\end{table}

\subsection{Continuous Integration}
Since several developers are working on the same project it is important to maintain a flawless \gls{CI} process. “Continuous integration is a development method where developers often incorporate code into a fully integrated system. Automated construction and automated tests can then check any integration.” \cite{mohsienuddin} This is ensured by using a GitLab server hosted by \gls{TH} running on a Docker container. \\

GitLab is an open DevOps platform for managing Git repositories ensuring version control. The key features contain “user-friendly web interface and the possibilities of managing permissions” \cite{hethey}.  Version control is a key factor in code development enabling developers to work simultaneously on the same project. The work can be saved at certain points and can be reverted to earlier stages without problems \cite{O'Grady}. \\

The \gls{CI} process consists of a pipeline with four stages: \texttt{install}, \texttt{quicktestData}, \texttt{quicktestSimulation}, and \texttt{styling}. These tests are automatically executed to ensure that the students, the simulations, and the visualization’s data are in a consistent form so they can be used by one another. \\ 

\begin{figure}[ht]
	\centering
	\small
  \includegraphics[width=1\linewidth]{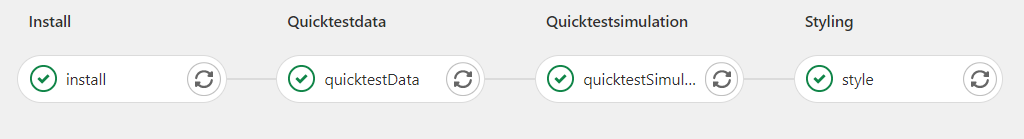}
	\caption{\texttt{The CI/CD pipeline implemented in GitLab consisting of four stages. By using a pipeline efficient development and trouble-free code integration was ensured.}}
	\label{fig:ci/cd}
\end{figure}

First, the install stage installs all libraries needed for the simulation. Next, the \texttt{quicktestData} stage tests if the \gls{CSV} file containing student’s data is read and transformed correctly into the format of choice by checking if the required results tables exist. \\
After checking the data format, the actual simulation is run. To ensure that the simulation was executed correctly, the \texttt{quicktestSimulation} stage checks if the simulation created new data, which can furthermore be analyzed and visualized. \\
Lastly, each file gets style checked by the R package \texttt{styler} \cite{styler} to maintain a clean code. If \texttt{styler} finds code style problems the user is asked to check his files locally and execute \texttt{stylers style\_file} function before committing new changes (see \textbf{Figure \ref{fig:ci/cd}}).

\subsection{Simulator}
\begin{figure}
	\centering
  \includegraphics[width=0.8\linewidth]{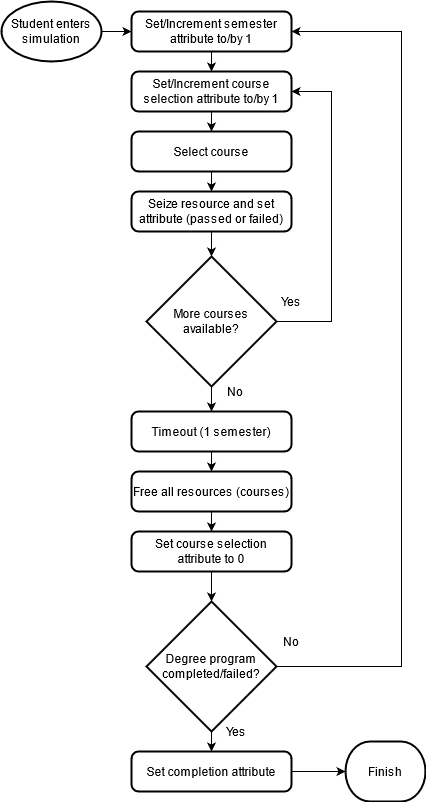}
	\caption{A simplified flowchart of the student's trajectory.\\ After setting the semester attribute and course selection counter to $1$ the student chooses a course through the \texttt{courseSelection} function. This process is then repeated \texttt{NUMBER\_COURSES} times, or till no more courses are available. Afterwards, the simulation advances by one semester and the whole process gets repeated until the student either completes all courses necessary for their program or leaves due to failing a course too often or by a random chance.}
	\label{fig:flow}
\end{figure}

For the implementation, \gls{simmer} a high-level process-oriented \gls{DES} package for \gls{R} was chosen \cite{Ucar19a}.
The chosen model has been heavily influenced by the work of Fiallos, Angel and Ochoa, Xavier \cite{8120908} which we discussed in section \ref{subsec:desrw}. A simplified flow chart of the simulation can be seen in \textbf{Figure \ref{fig:flow}}.
As mentioned earlier, students arriving in the simulation are assigned one of three groups depending on their average grade. This allows to consider different parameters, like failure rates in different exams for the groups, to get potentially more accurate results. The simulated students choose $n$ courses every semester based on their current state which they either complete or fail based on real-world probabilities calculated for each student group. \textbf{At the time of this writing the probabilities were chosen arbitrarily as no reliable data has been available.}\\ Students leave the simulation if they:
\begin{enumerate}[a)]
\item complete all courses that are part of their program.
\item fail a course $n$ times, where $n$ is a global constant that can be modified via the \gls{GUI}.
\item drop-out due to personal reasons implemented by a fixed probability each semester. This global constant can also be modified in the \gls{GUI}.
\end{enumerate}

 The course selection process is the most complex part of the simulation and takes the most computational resources. Unfortunately, \gls{simmer} does not allow a trajectory to seize multiple resources with one \texttt{seize()} command. Therefore, the function to select a module \texttt{courseSelection} has to be called $n$ times in order to select $n$ courses. Another option would have been to use the \texttt{clone()} function to split the trajectory into $n$ copies that each select and seize one course. Whilst that would be a more elegant solution it could not be implemented because currently there is no option in \gls{simmer} for clones to share attributes. Inside the function, the trajectory gains access to all courses, each with a base selection-likelihood. They are adjusted by currently five parameters which are added to the base selection-likelihood factor, if their conditions are fulfilled. The parameters are as follows:\begin{itemize}
\item[\texttt{SM\_1}:] The student's semester is equal to the intended semester for the course
\item[\texttt{SM\_2}:] The student's semester is greater than the intended semester for the course
\item[\texttt{SM\_3}:] The student's semester is at least two less than the intended semester for the course
\item[\texttt{SM\_4}:] Course is prerequisite for another course 
\item[\texttt{SM\_5}:] Course prerequisites are not fulfilled
\end{itemize}
Afterwards, the module with the highest adjusted selection-likelihood factor is chosen and the corresponding resource is seized. This process is then repeated $n$ times, where $n$ is the constant \texttt{NUMBER\_COURSES} determining how many courses each student chooses each semester. Currently, the selection is completely deterministic, meaning trajectories in the same state will always choose the same courses. It might be advisable to add a random factor to the likelihood to account for factors outside the scope of the simulation as it has been done in \cite{8120908}. The number of courses selected could also be changed to a random value which might differ depending on the student group, for this more data analysis has to be done. The whole selection process still needs refinement, as at the time of writing this report there was no reliable data available to optimize these parameters. The conditions were also chosen heuristically and should be modified or supplemented with additional conditions in the future according to need.  Most of these conditions have been implemented by simply comparing two vectors, or one vector with an attribute and only take reasonable computation time. Only \texttt{SM\_5}, the prerequisite check,  has been challenging to implement and is currently not a part of the prototype as in its current version it took too much computational time and slowed down the entire simulation fourfold. The function that checks whether the prerequisites are fulfilled, has to be applied to each available course of the program \texttt{NUMBER\_COURSES}-times for each student each semester. \\

\begin{figure}[ht]
	\centering
  \includegraphics[width=0.5\linewidth]{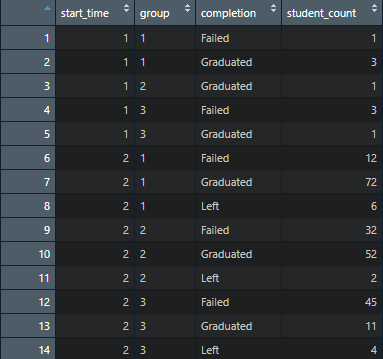}
	\caption{\texttt{dropout\_df}, one of three resulting data sets. The column completion shows the three reasons why the trajectories were left.}
	\label{fig:dropoutdf}
\end{figure}

After choosing a course and seizing the corresponding resource, the grading function according to their group is called. The student either passes or fails the course. Currently, the probabilities are chosen arbitrarily as no reliable data has been available to calculate the chance of succeeding in a given exam for each student group. If a student passes the course, the corresponding course attribute is increased by $1$. If they fail, it is increased by $0.1$. The latter has been chosen to indicate how often a student has failed a given course without having to introduce a new attribute. For example, if a student fails the course ''Mathematics 1'' three times and then passes in his fourth attempt the attribute \texttt{MA\_1} will be $ 0.1 + 0.1 + 0.1 + 1 = 1.3 $ at the end of the simulation. \\\\
Another idea which might be worth exploring, once data becomes available is that the likelihood of passing a course might change in consecutive attempts. If the student has attempted \texttt{NUMBER\_COURSES} many courses or has no available courses left the semester changes. In the context of the simulation this means a timeout of one time unit is applied and the student releases all resources. The simulation then checks if the student has completed all courses or failed one more often than allowed with \texttt{MAX\_ATTEMPTS}, or leaves because of the random \texttt{DROP\_OUT\_CHANCE}. If none of these apply, the process is repeated with the semester attribute being incremented by one. After the simulation is completed, the resulting data is used to create three data sets, that generate graphics for the associated Shiny app (see \textbf{Figure \ref{fig:dropoutdf}}).

\subsection{Visualizing Tools}
Simulation results mainly consist of integer numbers and certain string values paired for numerous rows which do not carry an appealing form at first sight. In order to communicate the generated outcomes with ease and leave less room for misinterpretation, visualization plays a crucial role. We built the \gls{GUI} based on Shiny, offering the users an interactive platform where they can adjust the simulation's parameters and receive the updated versions of graphs. This removes the barrier between men and machine one step further and helps them to interpret various scenarios with less effort. \\

Shiny is a web application framework and has two main components, \texttt{ui}  and \texttt{server}:
\begin{itemize}
\item \texttt{ui} is where the user interface of the application is defined. By default, Shiny utilizes the \gls{CSS} framework, Twitter Bootstrap. Yet it is possible to design the front end from scratch.
\item \texttt{server} defines the functionalities of the Shiny app, it connects the related part with each other. A Shiny app with a codeless server function will still run and deliver the created user interface, though it will be in an unresponsive state.
\end{itemize}

Reactivity is the key concept which enables the dynamic characteristic of a Shiny app \cite{masteringShiny}. The server logic is expressed by using reactive programming to automatically update the outputs when inputs change. Another important concept in Shiny is laziness. An app will only do the minimal amount of work needed to update the output avoiding unnecessary computation. This is accomplished by caching the outputs and the reactive connections. Shiny compiles the code into \gls{HTML}, CSS and JavaScript needed to display the application on the web.\\

Alongside Shiny, ggplot2 is used to create visualizations. ggplot2 is a widely used visualization tool for data analysis, its core lies on the principles of Grammar of Graphics \cite{gg}. Both Shiny and ggplot2 are R packages and available on \gls{CRAN}. \\

\subsection{Visualization Results}
The simulator provides seven different parameters. It allows the user to select the desired initial state to run the simulation.
\begin{enumerate}
    \item \textbf{Start Semester}: Defines the initial semester of simulation.
    \item \textbf{Start Year}: Defines the initial year of simulation.
    \item \textbf{End Semester}: Defines the final semester of simulation.
    \item \textbf{End Yea}r: Defines the final year of simulation.
    \item \textbf{Course Count per Semester}: Defines how many courses each student will take in each semester.
    \item \textbf{Maximum Failing Attempts}: Defines how many times a student can retake an exam.
    \item \textbf{Dropout Probability}: Defines the probability of a student leaving the study program.
\end{enumerate}

\begin{figure}[ht]
    \centering
    \includegraphics[width=0.4\linewidth]{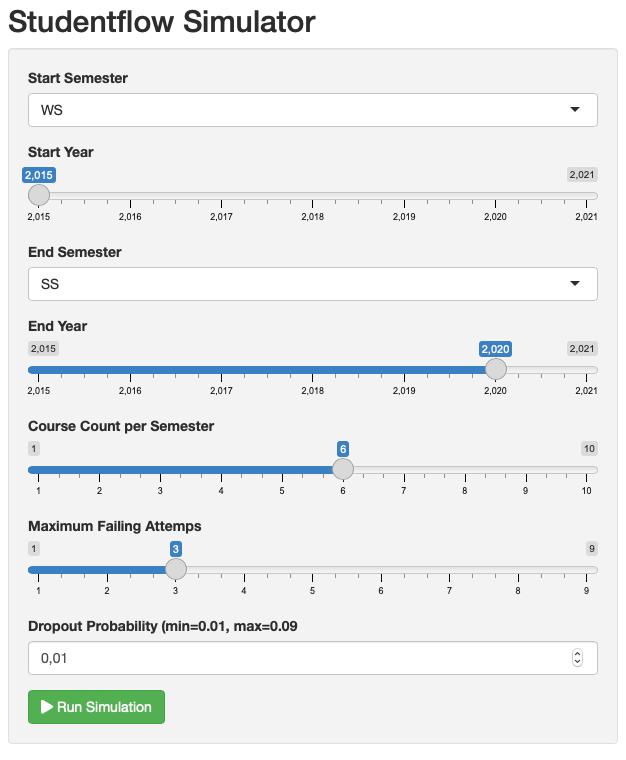}
    \caption{The Shiny app's sidebar panel}
\end{figure}

\begin{figure}[ht]
    \centering
    \includegraphics[width=1.1\linewidth,center]{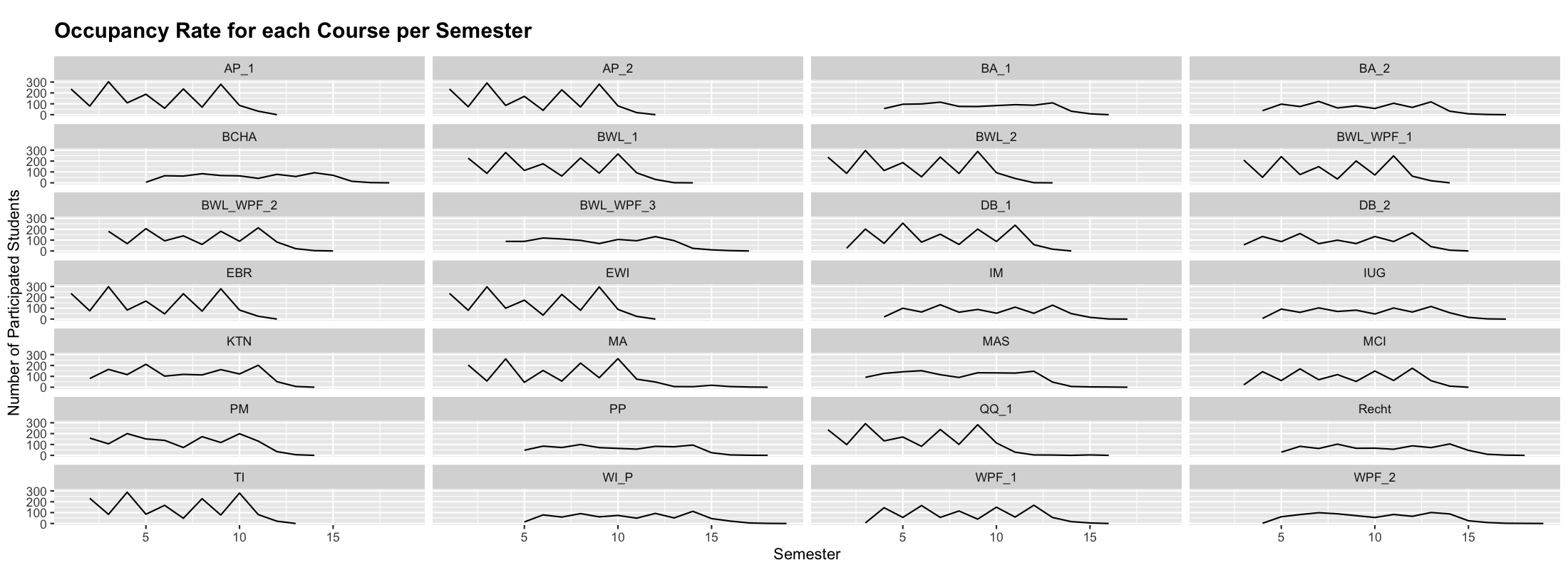}
    \caption{Course occupancy rate for each course and each semester.}
    \label{fig:occupancy-rate}
\end{figure}

After the simulation run is completed, the results are presented in the corresponding tab panels.
\newpage
\begin{itemize}
    \item \textbf{Graduation Rate}: illustrates the student flow for each student group and it depicts the completion state for each semester. A student either graduates, fails or leaves the study program. The groups are clustered depending on students exam results: 
    \begin{itemize}
        \item Group 1: Students who scored between 1.0 - 2.0 (very good) in their exams.
        \item Group 2: Students who scored between 2.0 - 3.0 (satisfactory) in their exams.
        \item Group 3: Students who scored between 3.0 - 4.0 (sufficient) in their exams.
    \end{itemize}

The number of students enrolling in the summer semesters is significantly lower. This occurs, because \gls{TH} mostly offers programs starting in winter semester (see \textbf{Figure \ref{fig:grad-rate}}).
    
    \begin{figure}[ht]
    \centering
    \includegraphics[width=1\linewidth,center]{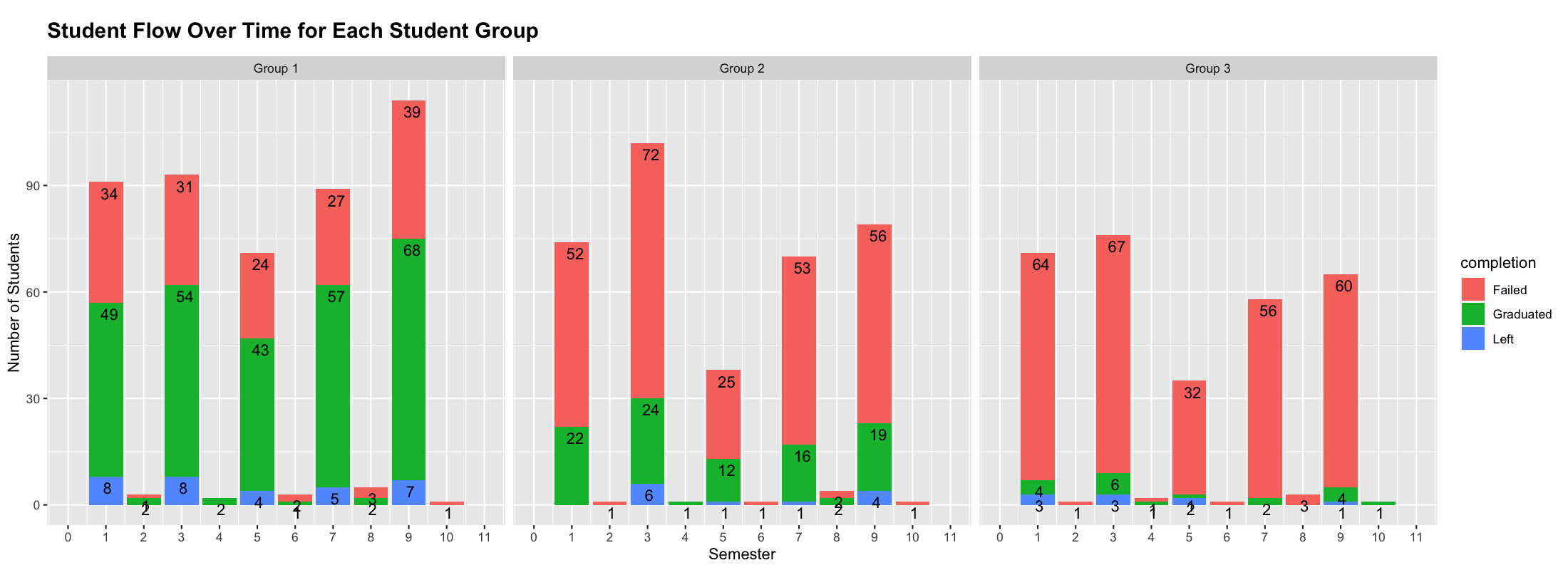}
    \caption{Graduation rate supplies a comparable view for student groups on a timeline, which makes it possible to interpret how well the students perform.}
    \label{fig:grad-rate}
\end{figure}

    \item \textbf{Study Duration}: This tab illustrates the average graduation time for each start semester and student's group. It serves as a direct comparison of student groups to evaluate their academic progress (see \textbf{Figure \ref{fig:grad-time}}).
    
    \begin{figure}[ht]
    \centering
    \includegraphics[width=1.1\linewidth,center]{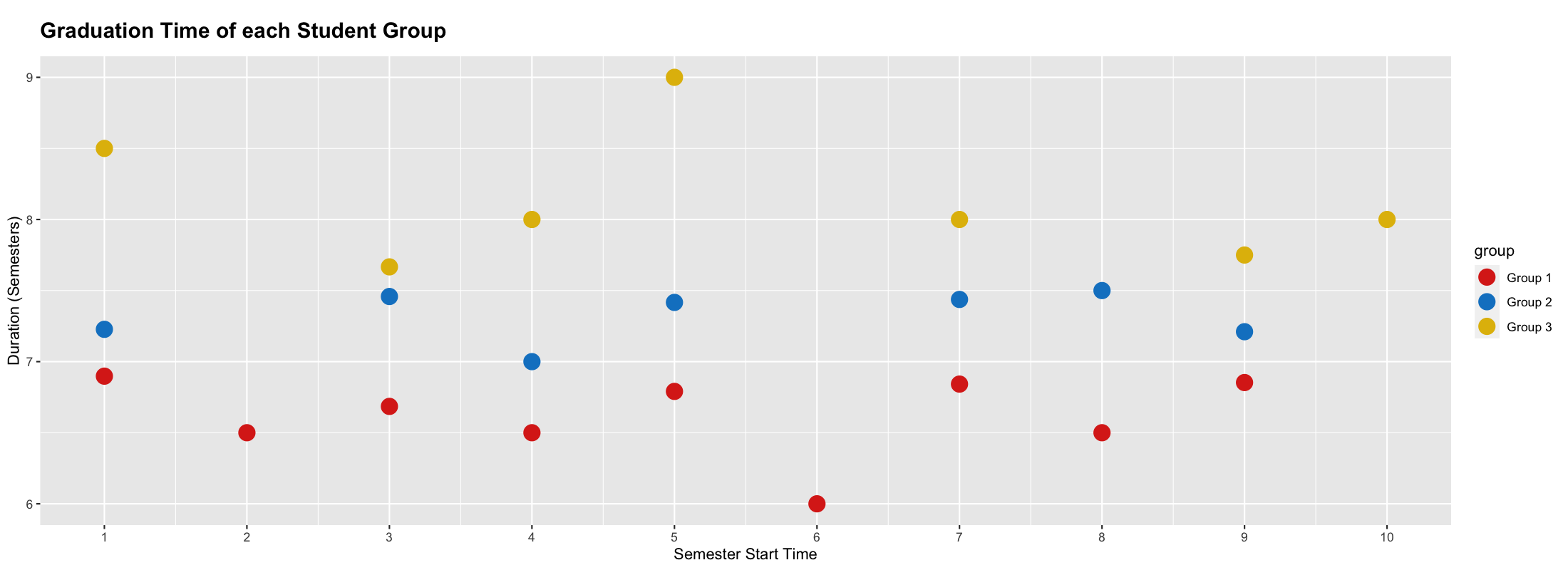}
    \caption{Study duration tab overview.}
    \label{fig:grad-time}
\end{figure}


    \item \textbf{Course Occupancy}: The last tab provides an overview about all courses offered and the number of students taking these courses each semester. This can help to detect bottlenecks caused by a specific course or semester (see \textbf{Figure \ref{fig:occupancy-rate}}).
\end{itemize} 

\section{Results}
Studentflow simulation tools offer an initial model designed to investigate the relationship between students and academic courses by projecting their academic performance and enrollment states. Visualizations demonstrate the significant differences between each group as well as a correlation between students with good grades and faster graduation time. 
Overall, we are confident that a discrete event simulation can be a promising tool to simulate student flows and should be taken into consideration by the university policymakers when designing academic programs. Moreover, universities can clearly spot out weaknesses in their current course design to actively act against higher dropout rates in certain classes.
\section{Discussion}
The biggest drawback of the simulation is the lack of real student data. In order to gain accurate results, the initial parameter values need to be tuned accordingly. The optimization of the course selection process will take a lot of exploratory data analysis to find out which are the most important factors in choosing a course. To achieve this, we propose gathering a decent amount of data to observe students' performances on a relatively long time scale. This would enable us to calculate the real success rate of students for each course and its occupancy rate. Instead of only using constants, one can introduce random factors to the parameter values, to get more dynamic results. The current model and its constraints are not sufficient yet to generate valuable insights. Therefore, the model has to be expanded. The improved model should consider the following aspects:
\begin{enumerate}[a)]
    \item Do multiple attempts improve the likelihood of passing an exam?
    \item How does the availability of a course in a given semester influence the attendance and the success rate for a chosen exam, which is offered at \gls{TH} every semester?
    \item What is the effect of prerequisite assignments to attend an exam? Can and should this be taken into account in the simulation?
    \item How should the course capacities be modeled? In reality, not all students attend to every lecture of a module they intend to complete. Because of this, discrepancies could occur between simulated occupancy and actual occupancy rates (i.e. students in a lecture).

\end{enumerate} 

\bibliographystyle{unsrt}
\newpage
\bibliography{bibliography.bib}
\end{document}